\documentclass[aps,prb,reprint,amsmath,amssymb,superscriptaddress,showpacs]{revtex4-2}
\usepackage{bm}
\usepackage{graphicx}
\usepackage[usenames, dvipsnames]{color}
\usepackage{dcolumn}
\usepackage{float}
\usepackage{hyperref}
\usepackage{tabularx}
\usepackage{siunitx}
\usepackage{xcolor}
\newcommand{\pt}{\color{black}} 
\newcommand{\bl}{\color{black}} 
\newcommand{\ch}{\color{black}}
\newcommand{\chk}{\color{black}}

\begin{document}

\title{Magnetocaloric properties of $R_3$Ga$_5$O$_{12}$ ($R$ = Tb,Gd,Nd,Dy)}
\author{M. Kleinhans}
\affiliation{Physik-Department, Technische Universit\"{a}t M\"{u}nchen, D-85748 Garching, Germany}

\author{K. Eibensteiner}
\affiliation{Physik-Department, Technische Universit\"{a}t M\"{u}nchen, D-85748 Garching, Germany}
\affiliation{kiutra GmbH, Rupert-Mayer-Str. 44, D-81379 M\"unchen, Germany}

\author{J. C. Leiner}
\email[]{Corresponding author: jon.leiner@tum.de}
\affiliation{Physik-Department, Technische Universit\"{a}t M\"{u}nchen, D-85748 Garching, Germany}
\affiliation{Heinz Maier-Leibnitz Zentrum (MLZ), Technische Universit\"at M\"unchen, D-85748 Garching, Germany}

\author{{\chk C. Resch}}
\affiliation{Physik-Department, Technische Universit\"{a}t M\"{u}nchen, D-85748 Garching, Germany}

\author{{\chk L.~Worch}}
\affiliation{Physik-Department, Technische Universit\"{a}t M\"{u}nchen, D-85748 Garching, Germany}

\author{{\chk M. A. Wilde}}
\affiliation{Physik-Department, Technische Universit\"{a}t M\"{u}nchen, D-85748 Garching, Germany}

\author{J. Spallek}
\affiliation{Physik-Department, Technische Universit\"{a}t M\"{u}nchen, D-85748 Garching, Germany}
\affiliation{kiutra GmbH, Rupert-Mayer-Str. 44, D-81379 M\"unchen, Germany}

\author{A. Regnat}
\affiliation{Physik-Department, Technische Universit\"{a}t M\"{u}nchen, D-85748 Garching, Germany}
\affiliation{kiutra GmbH, Rupert-Mayer-Str. 44, D-81379 M\"unchen, Germany}

\author{C. Pfleiderer}
\affiliation{Physik-Department, Technische Universit\"{a}t M\"{u}nchen, D-85748 Garching, Germany}
\affiliation{Center for Quantum Engineering (ZQE), Technische Universit\"at M\"unchen, D-85748 Garching, Germany}
\affiliation{Munich Center for Quantum Science and Technology (MCQST), Technische Universit\"at M\"unchen, D-85748 Garching, Germany}

\date{\today}

\begin{abstract}
We report the characteristic magnetic properties of several members of the rare-earth garnet family, Gd$_3$Ga$_5$O$_{12}$ (GGG), Dy$_3$Ga$_5$O$_{12}$ (DGG), Tb$_3$Ga$_5$O$_{12}$ (TGG), and Nd$_3$Ga$_5$O$_{12}$ (NGG), and compare their relative potential utility for magnetocaloric cooling, including their minimal adiabatic demagnetization refrigeration (ADR) temperatures and relative cooling parameters. {\pt A main objective of this work concerns the identification of potential improvements over the magnetocaloric properties of GGG for use in low temperature ADR cryostats. Using Tb$^{+3}$ and Dy$^{+3}$ in the rare-earth site offers, in principle, a higher saturation magnetization and Nd$^{+3}$ gives a lower \mbox{de Gennes} factor and therefore potentially reduced magnetic transition temperatures, limiting the useful temperature range. Our results show that Dy$_3$Ga$_5$O$_{12}$ yields an optimal relative cooling parameter ($RCP$) at low applied fields and low limiting temperatures, which would allow for the design of more efficient ADR cryostats. }   
\end{abstract}

\maketitle

\section{Introduction}
Adiabatic Demagnetization Refrigeration (ADR) techniques based on the magnetocaloric effect (MCE) are becoming an increasingly popular means of cryogenic cooling. Thus, materials that can further improve the performance of ADR devices are in high demand. 
The magnetocaloric effect is a phenomenon in which certain materials change temperature in response to an externally applied magnetic field. In essence, the temperature of such a material increases when an external magnetic field is applied in an adiabatic fashion and then, upon the removal of the applied field, adiabatic demagnetization (ADM) processes restore the original temperature and create a closed thermodynamic cycle. These thermodynamic responses and processes are typically highly reversible, and consequently exhibit energy efficiencies desirable in a variety of practical application contexts. In fact, since the interaction between the magnetic moments and the magnetic field coupling is a quantum mechanical effect, thermodynamic heat pumping cycles can approach 100\% efficiency when hysteresis and eddy currents are both negligible. 

Magnetic refrigeration technologies based on the MCE offer potential energy savings up to 30\%  \cite{alahmer2021magnetic} compared with other conventional techniques, and do not require the use of any refrigerant gases. Thus, the MCE solid-state technology has the potential to significantly reduce the environmental impact {\ch in comparison} to mainstream cooling technology. Such refrigeration technology is utilized in a diverse array of applications {\ch which include}: Magnetic Resonance Imaging (MRI) scanners, low temperature sensors (including detectors in satellites for astrophysical observations), spintronic devices, and in creating suitable environments for reliable quantum computing processes.


Precision categorization and quantification of both the universal and distinct ADM characteristics in the materials exhibiting the MCE is an ongoing endeavor. This knowledge would allow for {\ch the} tuning and optimization of these properties for use in the various potential applications outlined above \cite{WIKUS2011,WIKUS2014}. {\bl In this paper, we concentrate on materials with {\ch the} potential utility {\ch for} operating low-temperature cryostats between \SI{4.2}{K} and \SI{300}{mK}, {\ch such as} the ADR cryostats of kiutra GmbH \cite{kiutra}.}

\begin{figure}[t]
	\centering
	\includegraphics[width=1.005\columnwidth,clip]{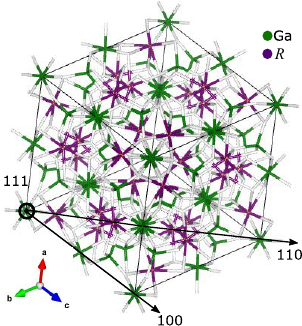}
	\caption{ \bl{Depiction of the garnet crystal structure $R_3$Ga$_5$O$_{12}$ seen along the \si{<}111\si{>} axis (space group Ia$\overline{3}$d, no. 230), with the relevant crystallographic directions marked. Broadly visible is the interlaced hyperkagome lattice motif consisting of cornersharing triangles with rare earths at the vertices, with an angle of \ang{70.5} between two neighboring triangles.} \cite{dunsiger2000low,petrenko1999magnetic} }
	\label{fig:garnet_structure}
\end{figure}

{\bl The following criteria must be considered in the search for an ideal refrigerant suited for low-temperature ADM applications: (i) large magnetic entropy densities which can be readily reduced through the MCE at low magnetic fields, (ii) no magnetic ordering {\ch within the operating temperature range}, (iii) low lattice entropy, (iv) near zero magnetic and thermal hysteresis, (v) high chemical stability, (vi) {\ch composition of noncorrosive substances}, (vii) straight-forward and cost efficient fabrication, (viii) high electrical resistance, (ix) {\ch high} thermal conductivity, and (x) {\ch high} specific heat.

Many of these criteria are fulfilled by magnetically frustrated systems, like those in the rare-earth garnet family. The different parameters such as compensation points on the magnetization curve, the saturation magnetization values, magnetocrystalline anisotropies, and lattice constants can be  adjusted by the substitution of different rare-earth elements, {\ch e.g.} Tb, Gd, Nd, and Dy, into the the garnet structure as shown in Fig. \ref{fig:garnet_structure}. This structure is geometrically frustrated in three dimensions by virtue of the interwoven hyperkagome lattice motif, {\ch which consists of} a two-dimensional (2D) network of corner-sharing triangles facilitating low magnetic ordering temperatures \cite{Yoshioka2004frustrated, PhysRevResearch.2.033509}. 

Geometrically frustrated materials are especially interesting for their nonhysteretic paramagnetic behavior down to {\ch  liquid helium temperatures}. {\ch Additionally, these materials have the desired attributes of} chemical stability, high electric resistance, noncorrosivity, and the possibility for scalable cost-efficient fabrication. This gives them considerable advantages compared to corrosive salts, such as iron aluminium alum (FAA) \cite{FAA,vilches1966measurements} already in use for ADR applications, as well as other metallic nonsuperconducting frustrated material systems like KBaYb(BO$_3$)$_2$ \cite{tokiwa2021frustrated, sanders2017magnetism,guo2019crystal,guo2019triangular} and YbPt$_2$Sn \cite{YPS,gruner2014unusual}, which exhibit better heat conduction properties but are more difficult and more expensive to fabricate. {\ch Other materials in the garnets family such as {G}d$_{3}${G}a$_{5-x}${F}e$_ {x}$O$_{12}$ \cite{mcmichael1993enhanced} and $R_{3}$({G}a$_{1-x}${F}e$_{x}$)$_{5}${O}$_{12}$ ($R$= {G}d, {D}y, {H}o; 0$<$ x$<$ 1) \cite{provenzano2003enhanced, numazawa2000magnetic} also showed enhanced MCE both in the sub-2-K region and as high as \SI{80}{K}. However, the complicated fabrication procedures of these crystals limits their growth into sufficiently large single crystals.}

For ADR applications in the \SI{0.3}{K} to \SI{4.2}{K} regime, materials which yield a large change in entropy $\Delta S$ in the temperature range of \SI{0.02}{K} to \SI{5}{K} are required. To optimize the ADR efficiency, it is desirable to have a strong magnetocaloric effect that manifests {\ch itself} within the lowest possible applied magnetic field value. This would significantly reduce the size and cost for the apparatus. The requirement of good thermal conductivity means that single-crystal samples are preferable. In general, polycrystalline materials have reduced thermal conductivity due to the influence of grain boundaries, which can only be partially remedied by increasing the packing density. 
}

\section{Outline}
\begin{figure}
	\centering
	\includegraphics[width=1.005\columnwidth,clip]{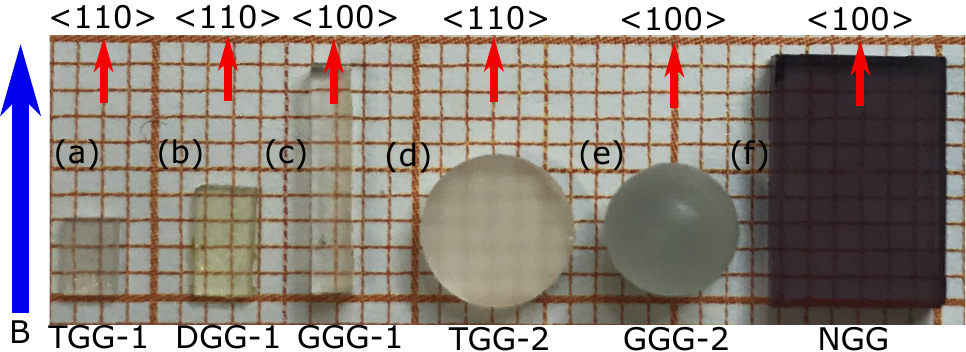}
	\caption{ Photograph of the samples used for the  measurements in this work, {\chk (Note: DGG-2 is not shown)}. The samples (a), (b), and (c) are used for VSM measurements in the PPMS. The larger samples (d), (e), and (f) are used for VCM measurements in the TL-400 for a better signal-to-noise ratio. The direction of the applied magnetic field $B$ is shown in blue. {\bl The measured crystallographic direction is indicated by the red arrows.}  }
	\label{Samples1}
\end{figure}

{\bl 

To determine the suitability for ADR of a given material between \SI{4.2}{K} and \SI{0.3}{K}, we mapped out the MCE (given as the value of the change in entropy $\Delta S(T,B)$) around the interesting point of operation (IPO) defined as $T$=\SI{4}{K} and $B$=\SI{3}{T}. These IPO values represent the typical starting conditions for an ADR system precooled by a pulse tube cryocooler. This map of $\Delta S(T,B)$ around the IPO is constructed from individual temperature scans under different applied magnetic fields, which for this study were performed between \SI{0.05}{K} and \SI{30}{K} {\ch and from zero} to \SI{6}{T}. 
$\Delta S(T,B)$ is extracted {\ch from integrating over} the $\frac{\partial M(T,B')}{\partial T}$ curves for several applied magnetic fields, {\ch which precisely quantifies} how much entropy may be released when a magnetic field is applied.

From the wide variety of materials in the rare-earth gallium-garnets category $R_3$Ga$_5$O$_{12}$ \cite{onn, Filippi_1980, Petrenko1998, Guillot_1985,cooke1967magnetic}, we narrow {\ch our focus down to} the four most promising candidates for ADR applications at T$<$ \SI{5}{K}. $R_3$Ga$_5$O$_{12}$ are classified by their common garnet crystalline structure where the magnetic ions, e.g. Gd, form two interpenetrating corner-sharing triangular sublattices. This structure yields a high degree of geometric frustration of the electron spins, which can lead to the absence of long-range magnetic order down to  $T$~$<$~1~{K}~\cite{ramirez1991}.

Gadolinium gallium garnet \cite{gschneidner_jr1999} (GGG) is a well-established MCE material for magnetic refrigeration and ADR systems in the \SI{4.2}-{K} temperature regime, due to the absence of long range ordering down to \SI{0.025}{K} at zero field. It exhibits several of the features which make it ideal for ADR systems: lack of long range order, lack of single ion anisotropy ($L = 0$ for Gd$^{3+}$), chemical stability, high electrical resistance, and a high density of magnetic ions with a large magnetic moment (Table \ref{tab:mce_data}). This allows for the full magnetic entropy (R$ln[2J + 1]$ = 17.29 J K$^{-1}$ mol Gd$^{-1}$) \cite{2017_Ln3CrGa4O12} to be obtained in magnetic fields under \SI{3}{T}. However, GGG develops short-range order at temperatures below \SI{0.8}{K} and a small thermal hysteresis. This limits its suitability below \SI{1}{K} because the full $\Delta S$ effectively cannot be accessed. 

 To find a material with a higher $\Delta S$ at the IPO {\ch as well as suppressed magnetic ordering temperature}, we compare other rare earth garnets with GGG. The three additional samples in this category which are characterized in our study, are single crystals of $\mathrm{Nd_3Ga_5O_{12}}$ neodymium gallium garnet (NGG) \cite{onn,hamman:jpa-00207421,antic1986crystal}, $\mathrm{Tb_3Ga_5O_{12}}$ terbium gallium garnet (TGG) \cite{wawrzynczak2019magnetic,gruber2007analyses,low2014magneto,low2013magnetization,hammann1975hyperfine} and $\mathrm{Dy_3Ga_5O_{12}}$ dysprosium gallium garnet (DGG).\cite{filippi1977specific,tomokiyo1985specific,heinz1972properties,goshorn1977specific, matsumoto, kimura, PhysRevResearch.2.033509} The sample shapes are shown in Fig. \ref{Samples1} and listed in Table \ref{tab:samples}. NGG and TGG are used as laser crystals, and so are readily available as low-cost high-quality single crystals from a number of suppliers. Additionally, TGG is shown to exhibit the phonon Hall effect, which is known to arise from magnetic ions coupling to lattice excitations.\cite{PRL_PhononHallEffect_TGG, inyushkin2007phonon} DGG, while a promising candidate for use in ADR systems generally, is not easily grown in the form of large enough single crystals for the ADR systems considered in this study. 

Other reasons for examining the {\ch aforementioned} materials are their single-ion properties, as outlined in Table \ref{tab:mce_data}, which are close to {\ch those of} GGG. Table \ref{tab:mce_data} shows the {\ch respective} values of the Lande factor, g$_J$, the azimuthal quantum number, $J$, the theoretical maximum saturation magnetization, g$_{J}J$, and the \mbox{de Gennes} factor, $\xi = (g-1)^2 J(J+1)$. The \mbox{de Gennes} factor is often proportional to the Curie temperature of a ferromagnetic material, but can also {\ch account for} the ordering temperatures of frustrated systems like the garnets. \cite{sharoyan2007dependence} As may be seen that Dy$^{3+}$ and Tb$^{3+}$ are {\ch especially promising} candidates for replacing Gd$^{3+}$, because, in principle, they could combine a higher saturation magnetization $M_{\mathrm{sat}}$ with a similar Lande factor g$_{J}$. With Dy$^{3+}$ and Tb$^{3+}$, there is a lower \mbox{de Gennes} factor than for Gd$^{3+}$, which indicates a lower phase transition temperature and hence a better minimal ADR temperature T$_{\mathrm{min}}$ (i.e. the minimum temperature attainable by demagnetization). Nd$^{3+}$ {\ch yields} the lowest \mbox{de Gennes} factor out of the four materials, although it yields a much lower saturation magnetization compared to Gd$^{3+}$. \cite{wawrzynczak2019magnetic,low2013magnetization,nekvasil1974anisotropy} Relevant atomic values of the rare-earth elements are summarized along with the main results of this study in Table \ref{tab:mce_data}.}

\begin{table*}[ht!]
    	\centering
        \begin{tabular}{|c|c|c|c|c|c|c|c|c|}
        	\hline
        	Material & Lattice Constant  & Easy Axis & Density &   Molar Mass  & Mass & Shape & Dimensions & Orientation  \\
        	           &(\AA) &  & \si{g/cm^3} & \si{g/mol}  &   g       & &   mm   &   \\ \hline
        	GGG-1  & 12.382 & 100  &   7.08    &   1012       &   0.1258 & cuboid &   1.48 x 1.46 x 8.48 & [100]x[010]x[001]\\
        	{\chk GGG-2} & 12.382 & 100 & 7.08 & 1012 & 0.40910 & sphere &   \o 4.8 & --- \\ 
        	DGG-1 & 12.307 & 111 &   7.30  &    1028     &  0.0305 & platelet  &   2.1 x 4.1 x 0.5 & [110]x[110]x[001] \\
        	DGG-2 & 12.307 & 111 & 7.30 & 1028 & 0.00750 & cube & 1.0 x 1.0 x 1.0 & [110]x[011]x[100] \\
        	TGG-1     & 12.355 & 111    &   7.13    &   1017       &   0.0186  & platelet  &   2 x 2.9 x 0.48 & [211]x[110]x[111]  \\
        	TGG-2 & 12.355 & 111 & 7.13 & 1017 & 0.3091 & disk & 5.58 x 5.58 x 1.77 & [110]x[211]x[111]\\
        	NGG    & 12.512 & 100 & 7.09   &  973 &  0.7218 & cuboid & 6.2 x 9.14 x 2 & [100]x[010]x[001]\\	\hline
        \end{tabular}
    	\caption{ {\bl Properties, shapes, dimensions, and orientations of the garnet materials studied in this work.}}
    	\label{tab:samples}
\end{table*}

\section{Methods}

\subsection{Sample Preparation}
{\bl All of the $R_3$Ga$_5$O$_{12}$ materials investigated in this study form in the same Ia$\overline{3}$d cubic crystal structure, illustrated in Fig.~\ref{fig:garnet_structure}. Each of the samples studied, as well as their measured crystallographic orientations, are shown in Fig.~\ref{Samples1}. NGG and GGG are measured with the magnetic field applied in the $[100]$ direction, {\ch while} TGG and DGG are measured {\ch with $B$} along the $[110]$ axis. The orientations are determined by Laue x-ray diffraction, and the samples are mounted accordingly for measurements along {\ch the different dimensions}.

All of the samples are supplied as cylinders. The DGG, NGG, and GGG crystals are {\ch obtained} from IMPEX Hightech, {\ch whereas the} TGG {\ch is purchased} from Electro-Optics Technology. A cuboid is cut from the material of each sample, as {\ch displayed} by the samples labeled TGG-1, {\ch DGG-1}, GGG-1 and NGG in Fig.~\ref{Samples1}. The material properties and dimensions are listed in full detail in Table \ref{tab:samples}. The samples are cut along the growth direction, i.e. along one of the three crystallographic directions, since this direction can be used in ADR systems without explicitly determining it and having to make {\ch specific} cuts of the single crystals. Additionally, for measurements at $T <$~3~{K}, a disk TGG-2, {\ch a sphere GGG-2,} {\chk and a cube DGG-2} are prepared.} 

\subsection{Experimental apparatus}
The magnetization data over a temperature range spanning from \SI{0.05}{K} to \SI{300}{K} is obtained using {\ch four} different systems. An Oxford Instruments vibrating sample magnetometer (VSM) \cite{foner1956vibrating} and a standard Quantum Design \SI{14}{T} PPMS with a vibrating sample option are used to carry out measurements from 3 up to \SI{300}{K}. A \SI{5}{T} $^3$He/$^4$He dilution refrigerator from Oxford Instruments, the TL-400, with a custom built vibrating-coil-magnetometer (VCM) \cite{smith1956development, maksimochkin2003VCM} is used to record magnetization data in the temperature range between \SI{50}{mK} and \SI{3}{K}. {\chk Additionally, a membrane based Faraday magnetometer (FMM) inside a \SI{15}{T} cryostat with a gradient coil is used together with a $^3$He insert for measurements between \SI{0.3}{K} and \SI{1.5}{K}}. 

The VCM is similar to a VSM, though in a VCM the detection coils vibrate in order permit cooling the sample \cite{VCM_technical_paper}. Absolute values of the magnetization of a sample are calculated from the VSM measurements with a calibration factor obtained by measuring a thin sheet of Ni as reference sample with a known magnetic moment.

The low temperature data recorded with the VCM is adjusted to fit the VSM data at \SI{3}{K}. For the VCM measurements, a field-dependent background is subtracted from the magnetization data. Measurements are carried out in fields of up to \SI{9}{T} in the (VSM) helium-flow cryostat and up to \SI{5}{T} in the (VCM) dilution refrigerator.

{\chk The Faraday magnetometer is custom built using a \SI{200}{nm} SiNx membrane coated with Ti/Au. The deflection of the membrane due to the magnetic force acting on the sample is measured capacitatively. Descriptions of similar setups may be found in the literature \cite{matsumoto2000construction, swanson1990diaphragm}. Absolute values of the magnetization are calibrated first by applying a voltage between the coated membrane and the counter electrode. This method is complimented in terms of a comparison to the VSM data recorded at high temperatures. Demagnetization effects due to different sample geometries for different sample shapes are taken into account. Measurements are performed for fields up to \SI{1}{T} for a magnetic gradient field of \SI{0.943}{T/m}. Every temperature scan is repeated with a negative gradient field in order to separate torque and force components of the signal.}
\subsection{Measurement conditions}
{\bl The samples are installed with the long axis parallel to the applied field in order to minimize the effects of demagnetizing fields. Two TGG and GGG samples are used: the small TGG-1 plate and GGG-1 cuboid for the VSM measurements and the TGG-2 disc and GGG-2 sphere for the VCM measurements (a larger sample in the VCM gives a better signal-to-noise ratio). {\ch In order to reduce the error in the VSM measurements due to the larger sample, separate larger calibration nickel samples are used. With these, systematic errors could be reduced to below 5\%.} All samples are glued to the sample holder with GE varnish for good thermal contact.}
{\chk The sample cube DGG-2 is prepared for measurements on the FMM by gluing it on the membrane with Stycast 2850 FT 24V instead of GE varnish. The Stycast provided higher mechanical stability preventing possible small positional changes of the sample, however, at the expense of reduced thermal coupling. The sample is oriented the same way as with DGG-1; the [110] axis parallel to the external field.} {\chk All} temperature scans are obtained using a field-heated - field-cooled protocol (FH-FC), when the sample is first warmed up in the presence of a constant applied field and then cooled down again under the same field.

\subsection{Entropy and cooling power}
 The magnetocaloric effect follows from the Maxwell relation linking the magnetization $M$ of a material with its entropy $S$ in terms of derivatives with respect to the magnetic field $B$ and the temperature $T$~\cite{magnetocaloric_effect, pecharsky1999magnetocaloric}. $\Delta S(T,B)$ may then be extracted from the magnetization as a function of applied magnetic field by virtue of an integration over $B$ at a given temperature $T$\cite{de2010theoretical}:
  \begin{equation} \label{eq:mce_calc}
        \Delta S(T,B) = \frac{1}{\mu_0} \int \limits_0^B \left ( \frac{\partial M(T,B')}{\partial T} \right )_{B,p} dB'
    \end{equation}

The temperature dependence of $\partial M(T,B') / {\partial T}$ at different fields is interpolated from zero field to the largest field measured. The integration of these curves with respect to the magnetic field permitted determination of the MCE, $\Delta S(T,B)$.

A metric generally used to characterize refrigeration materials is the refrigerant capacity (RC) \cite{wood1985general}, or, as otherwise known, the relative cooling power ($RCP$). The $RCP$ is defined as

\begin{equation} \label{eq:RC(H)}
        S_{RCP}(B) = \frac{1}{\mu_0} \int \limits_{T_{i}}^{T_{f}}  \Delta S(T,B) dB'
\end{equation}
where $T_i$ and $T_f$ are the temperatures of the reservoir at the beginning and at the end of a cycle, {\ch respectively}. Since a paramagnetic material often {\ch exhibits} a peak in its magnetocaloric effect, the area under the curve is frequently approximated as the product of the peak of the change of entropy. $\Delta S(T,B)$, {\ch multiplied by} FWHM of the peak of the MCE:
\begin{equation} \label{eq:RCP}
    S_{RCP}= - \Delta S_{\mathrm{max}} \delta T_{\mathrm{FWHM}}
\end{equation}
where $ \delta T_\mathrm{FWHM}$ is the temperature spanned by {\ch the FWHM of the} MCE peak and $ \Delta S_{max} $ the maximum value of the MCE peak at a given applied magnetic field. \cite{franco2018magnetocaloric,gschneidner_jr1999,provenzano2004reduction}

Another key metric for an ADR refrigerant is the minimum attainable temperature by demagnetization, $T_\mathrm{min}$, which {\ch is frequently limited by} a magnetic or structural phase transition. 
\\
\\
\section{Results and Discussion}
\begin{figure}[ht!]
        \centering
        \includegraphics[width=1.05\columnwidth,clip]{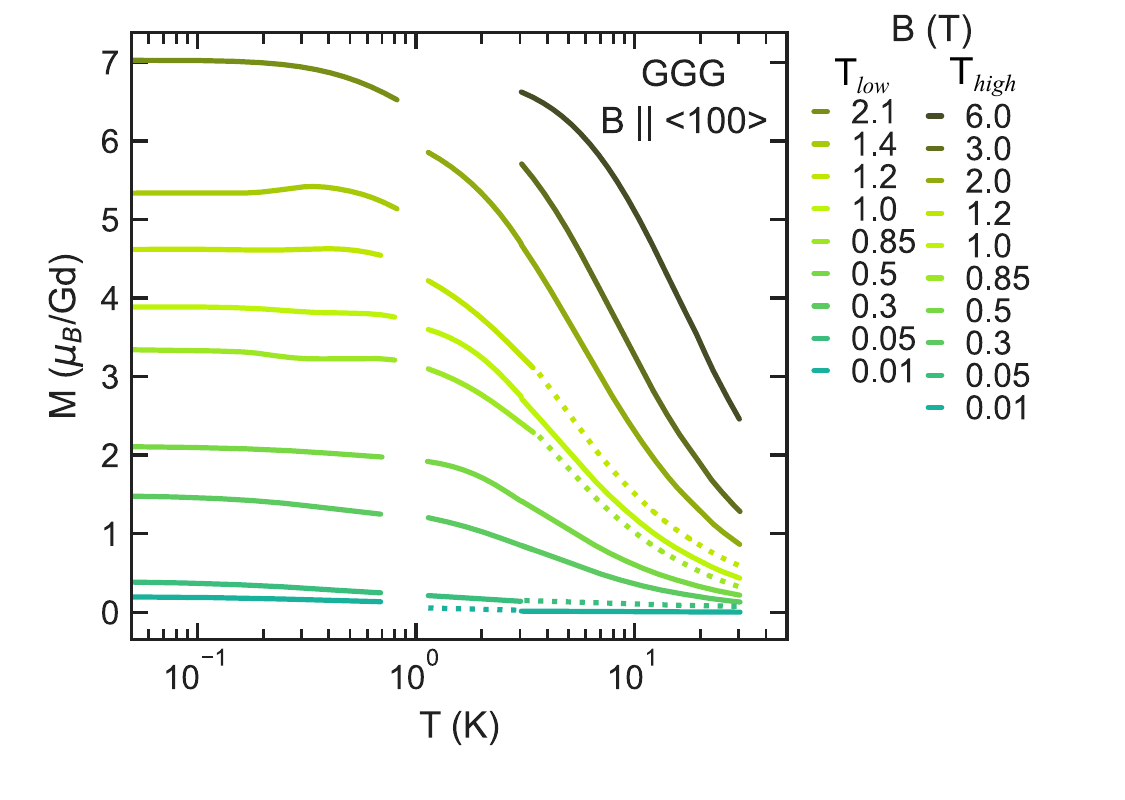}
        \caption{Magnetization of Gd$_3$Ga$_5$O$_{12}$ as a function of temperature. Data is recorded in the temperature range from \SI{60}{mK} to \SI{30}{K} for fields of \SI{10}{mT} to \SI{6}{T}. {\bl For the $T_{low}$ regime, \SI{60}{mK} to \SI{1}{K}, the GGG-2 spherical sample is used. For the $T_{high}$ regime between 1.3 to \SI{30}{K}, the GGG-1 cuboid is used. The {\ch dashed} lines represent extrapolations. } }
        \label{fig:ggg_mag}

	\centering
	\includegraphics[width=1.05\columnwidth,clip]{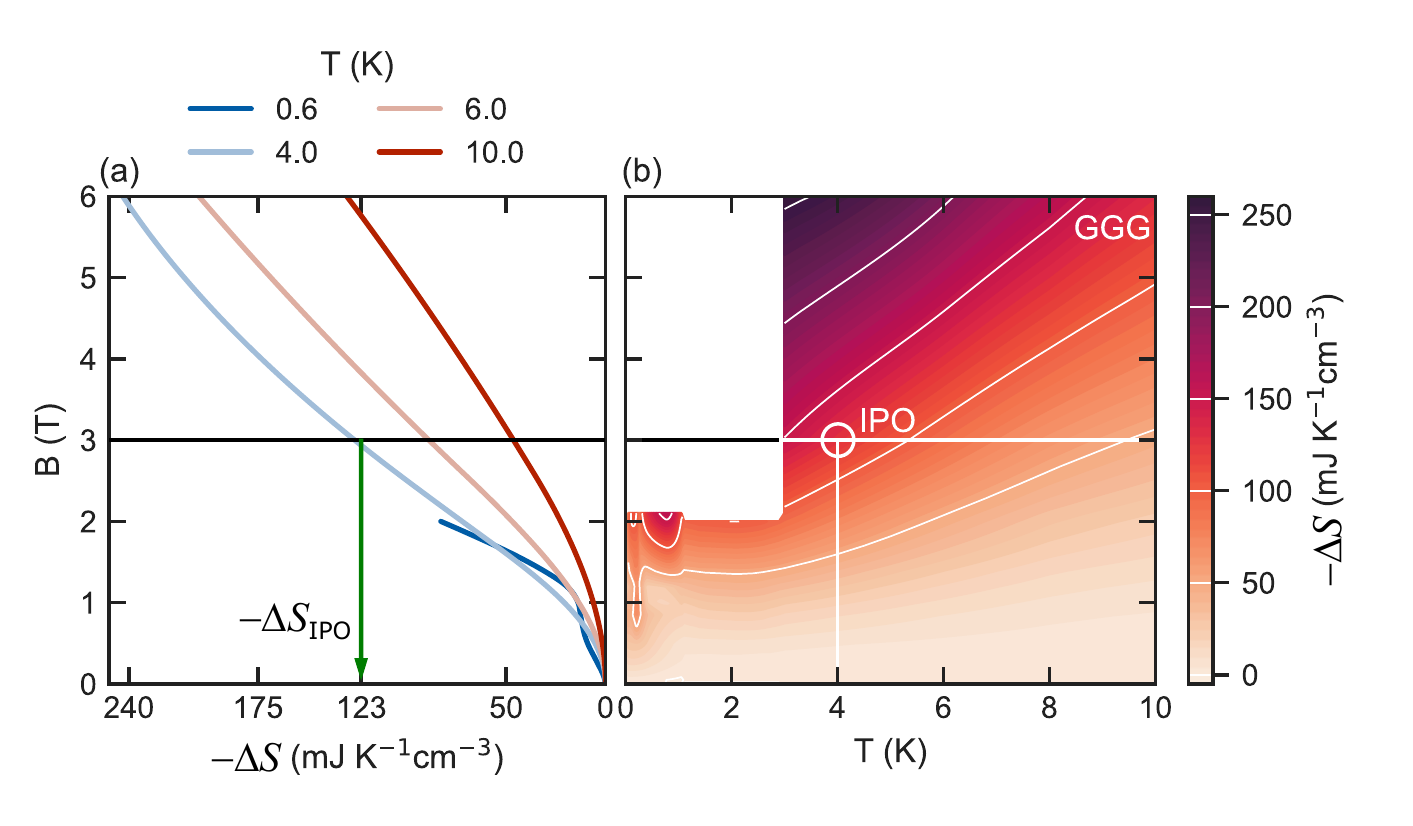}
	\caption{ Figure of merit for the magnetocaloric effect in Gd$_3$Ga$_5$O$_{12}$. Panel (a) shows the field dependence for each temperature measured. { \bl The value of the MCE at the interesting point of operation (IPO) at T=\SI{4}{K} and B=\SI{3}{T} is marked with a green {\ch arrow}. Panel (b) shows a contour plot of the MCE as a function of temperature and magnetic field, where a white circle marking the IPO.} At approximately \SI{300}{mK} and \SI{1}{T}, GGG enters its antiferromagnetic phase. This plot for GGG as well as the plots for DGG, NGG, and TGG shown in Figs. \ref{fig:dgg_contour}, \ref{fig:ngg_contour} and \ref{fig:tgg_contour}, respectively, correspond in style to Ref. \cite{2017_Ln3CrGa4O12}.}
	\label{fig:ggg_contour}
\end{figure}

\subsection{Gd$_3$Ga$_5$O$_{12}$ (GGG)}


{\bl The $RCP$ of GGG is calculated using typical temperature scans between \SI{0.025}{K} and \SI{300}{K} for different magnetic fields up to \SI{6}{T}, as shown in Fig. \ref{fig:ggg_mag}.} GGG shows a nonhysteretic increase in magnetization with decreasing temperatures featuring a Curie-Weiss dependence as expected of a paramagnetic material. From these data, the magnetocaloric effect is extracted and shown in Fig.~\ref{fig:ggg_contour}. The data below \SI{3}{K} is obtained from measurements with the spherical sample shown in Fig.~\ref{Samples1}. 

{\bl At low temperatures (T $<$ \SI{0.5}{K}) and magnetic fields ($B$ $>$ \SI{2}{T}), GGG reaches its saturation magnetization of $M_{\mathrm{sat}} = 7.1 \si{\micro _B/R}$ \cite{Hu_2014}. GGG shows a large MCE value of over \SI{250}{mJ~K^{-1}cm^{-3}} at \SI{6}{K} down to less than \SI{2}{K} and an MCE value of \SI{122}{mJ~K^{-1}cm^{-3}} at the IPO of \SI{4}{K} and \SI{3}{T}. This large {\ch MCE} value together with a large $S_{RCP}$ of \SI{2320}{mJ/cm^{3}} {\ch means that GGG is a benchmark} for MCE materials used in ADR cryostat stages at temperatures below\SI{20}{K} \cite{hov1980magnetic,hornung1974magnetothermodynamics,barclay1982materials,ancliff2021model,rousseau2017anisotropic,paddison2015hidden}}. 

{\ch However, }one issue with GGG that is not fully visible in the MCE plot is {\ch the sharp} MCE decrease between \SI{1}{K} and \SI{0.8}{K}. Here GGG enters a phase with short-range correlations despite still being paramagnetic. This may be seen in the broad peak in the heat capacity, which causes the GGG to loose a large portion of its cooling power when approaching this temperature range, {\ch as documented previously} \cite{Deen,Schiffer,Petrenko}.

{\pt We now compare the other commercially available rare-earth garnets with GGG} regarding their suitability as ADR refrigerants, searching specifically for materials that have good MCE properties below 0.8 K while not being prohibitively expensive. Some Gd containing MCE materials are previously reported to have better performances at around \SI{2}{K} \cite{lorusso2013dense,chen2014study,palacios2014magnetic,chen2015brilliant}.

\subsection{Dy$_3$Ga$_5$O$_{12}$ (DGG)}
Dysprosium gallium garnet, DGG, is known to be a promising replacement for GGG in ADR systems \cite{kimura} because it provides {\ch exceptional} cooling power at {\ch lower fields and} temperatures ($<$~\SI{0.5}{K}) \cite{filippi1977specific}. However, growing large single crystals is difficult, limiting its commercial production as a result. {\pt DGG has a large g$_{J}$ value of $4 / 3$, which is a significant fraction of the value for GGG g$_{J}$=2 \cite{WIKUS2014}, thus corresponding to large values of $\Delta S(T,B)$.}

\begin{figure}[]
        \centering
        \includegraphics[width=1.05\columnwidth,clip]{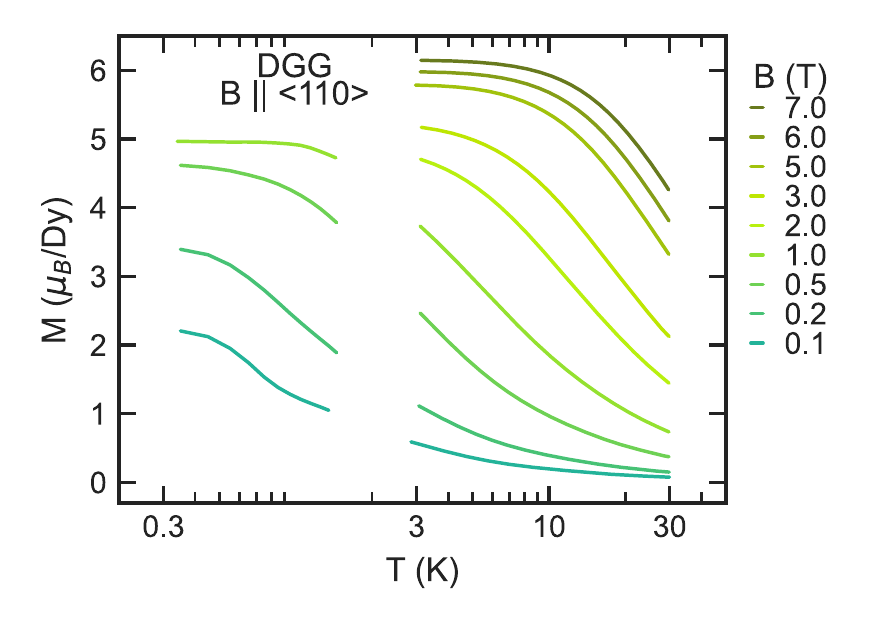}
        \caption{Magnetization of Dy$_3$Ga$_5$O$_{12}$ as a function of temperature. Data is recorded in the temperature range from 0.33 to \SI{30}{K} for fields of \SI{100}{mT} to \SI{7}{T}.}
        \label{fig:dgg_mag}
\end{figure}

\begin{figure}[]
	\centering
	\includegraphics[width=1.05\columnwidth,clip]{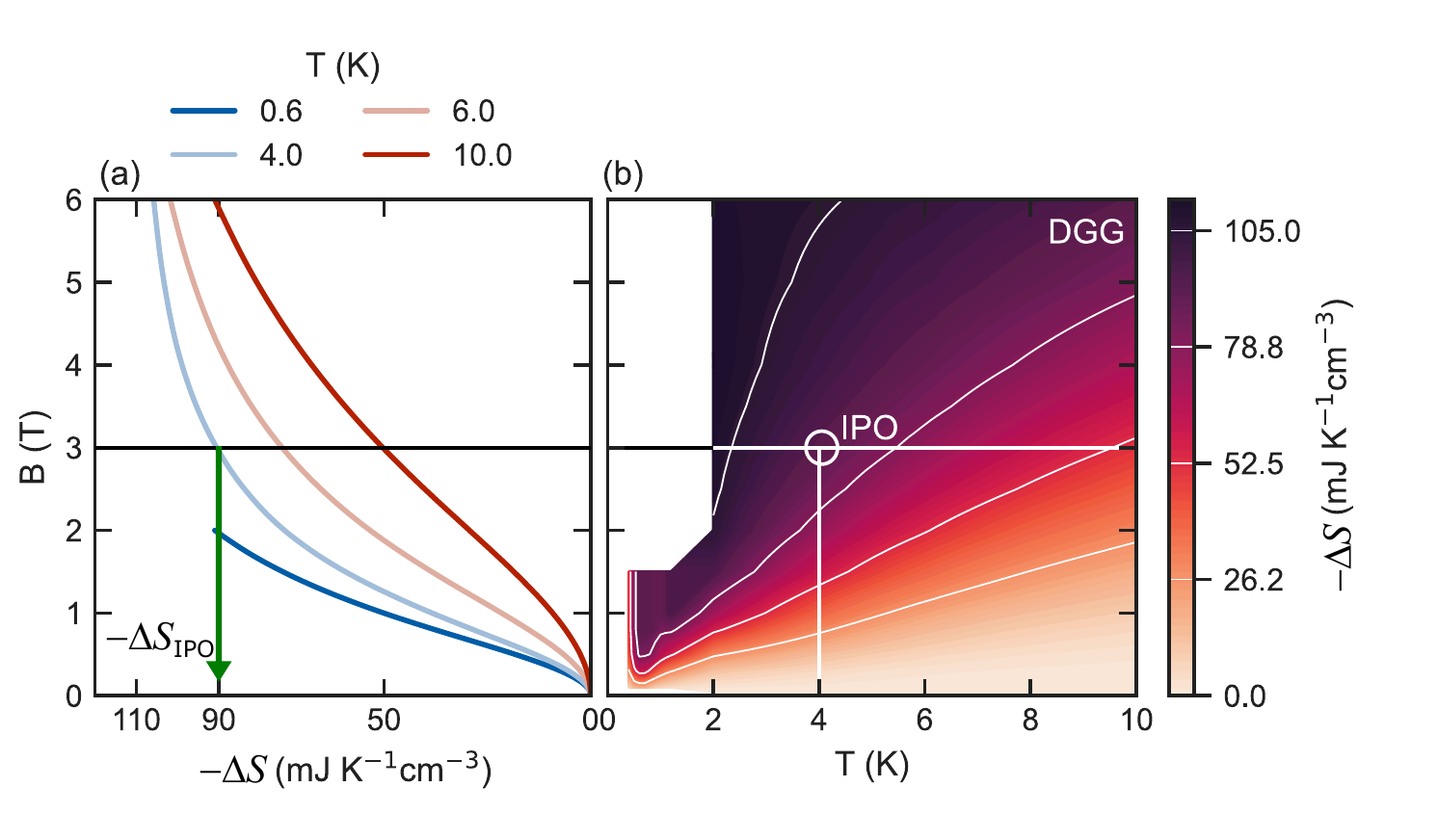}
	\caption{ Figure of merit for the magnetocaloric effect in Dy$_3$Ga$_5$O$_{12}$. This material is saturated around \SI{2}{T} at \SI{2}{K}, where it reaches its maximum MCE value {\ch of} $\Delta S$~=~\SI{106}{mJ~K^{-1}cm^{-3}} (Table~\ref{tab:mce_data}) with higher fields showing no effect on the entropy. { \bl The MCE value of \SI{90}{mJ~K^{-1}cm^{-3}} at \SI{3}{T} and \SI{4}{K} (IPO) is indicated in both panels.}}
	\label{fig:dgg_contour}
\end{figure}

Experiments show that {\pt DGG yields slightly better MCE {\ch values} in the temperature range of \SI{1.8}{K} to \SI{15}{K} {\ch when compared to} GGG, without a large peak in the heat capacity reducing its cooling capacity \cite{kimura, matsumoto},} while $M_{\mathrm{sat}} = 6.2 \si{\micro _B/RE}$ ion is reduced in comparison to GGG ($M_{\mathrm{sat}} = 7.1 \si{\micro _B/RE}$ ion) as shown in Table \ref{tab:mce_data}. 

{ \bl With a $\Delta S_{\mathrm{max}}$ of \SI{106}{mJ~K^{-1}cm^{-3}}, DGG is located between the values of $\Delta S_{\mathrm{max}}$ observed for NGG and GGG. Our results show that DGG is able to reach {\ch most of} this high value of $\Delta S_{\mathrm{max}}$  even at the IPO of \SI{4}{K}/\SI{3}{T}}. The magnetization of DGG between \SI{30}{K} down to \SI{0.33}{K} is shown in Fig. \ref{fig:dgg_mag}. DGG reaches its maximum saturation magnetization at around \SI{3}{K} and \SI{7}{T}. These data are used to calculate the MCE shown in Fig. \ref{fig:dgg_contour}. The magnetocaloric effect is saturated at \SI{2}{K} and \SI{2}{T}. DGG is known from previous work to have a antiferromagnetic phase transition at \SI{373}{mK} {\ch for zero field} \cite{filippi1977specific}. {\chk In our measurements with a minimum field of \SI{0.1}{T}, no transition is found. Possible reasons for this include a shift of the transition to lower temperatures under applied fields, or that the sample itself is slightly warmer than the sample thermometer. For the purpose of this study, the highest transition temperature of \SI{0.37}{K} at zero field is assumed as the lowest possible temperature for adiabatic demagnetisation. With this assumption, DGG reaches a large RCP of approximately $S_{RCP}$ = \SI{940}{mJ/cm^{3}}. } \\

 \begin{figure}[t]
        \centering
        \includegraphics[width=1.05\columnwidth,clip]{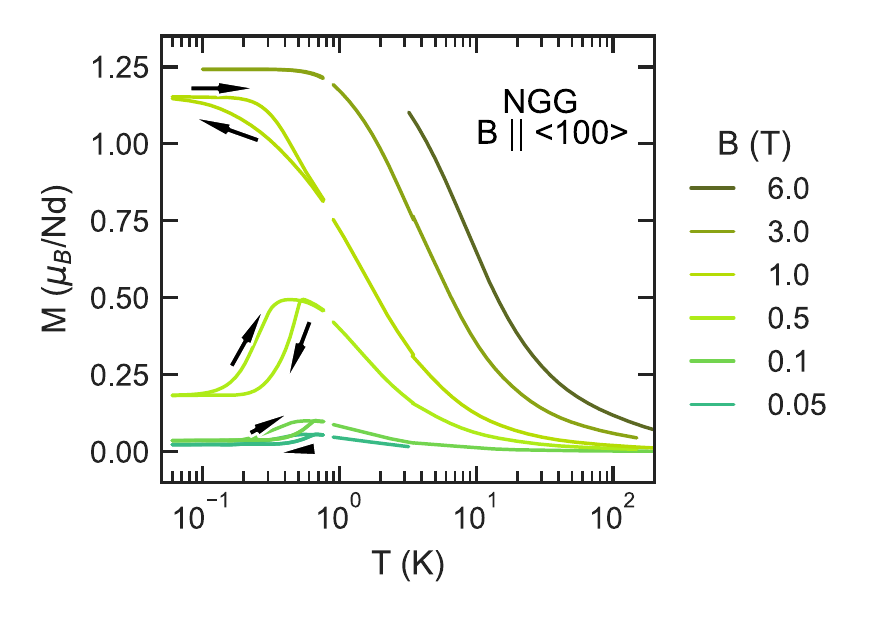}
        \caption{Overview of  temperature scans for NGG obtained in the VSM and VCM. The magnetization per neodymium ion is plotted against temperature. The magnetic field is applied along the [100] direction. {\bl The arrows indicate the sweep direction in the hysteresis loops under \SI{0.6}{K}.} }
        \label{fig:ngg_mag_fullrange}

	\centering
	\includegraphics[width=1.05\columnwidth,clip]{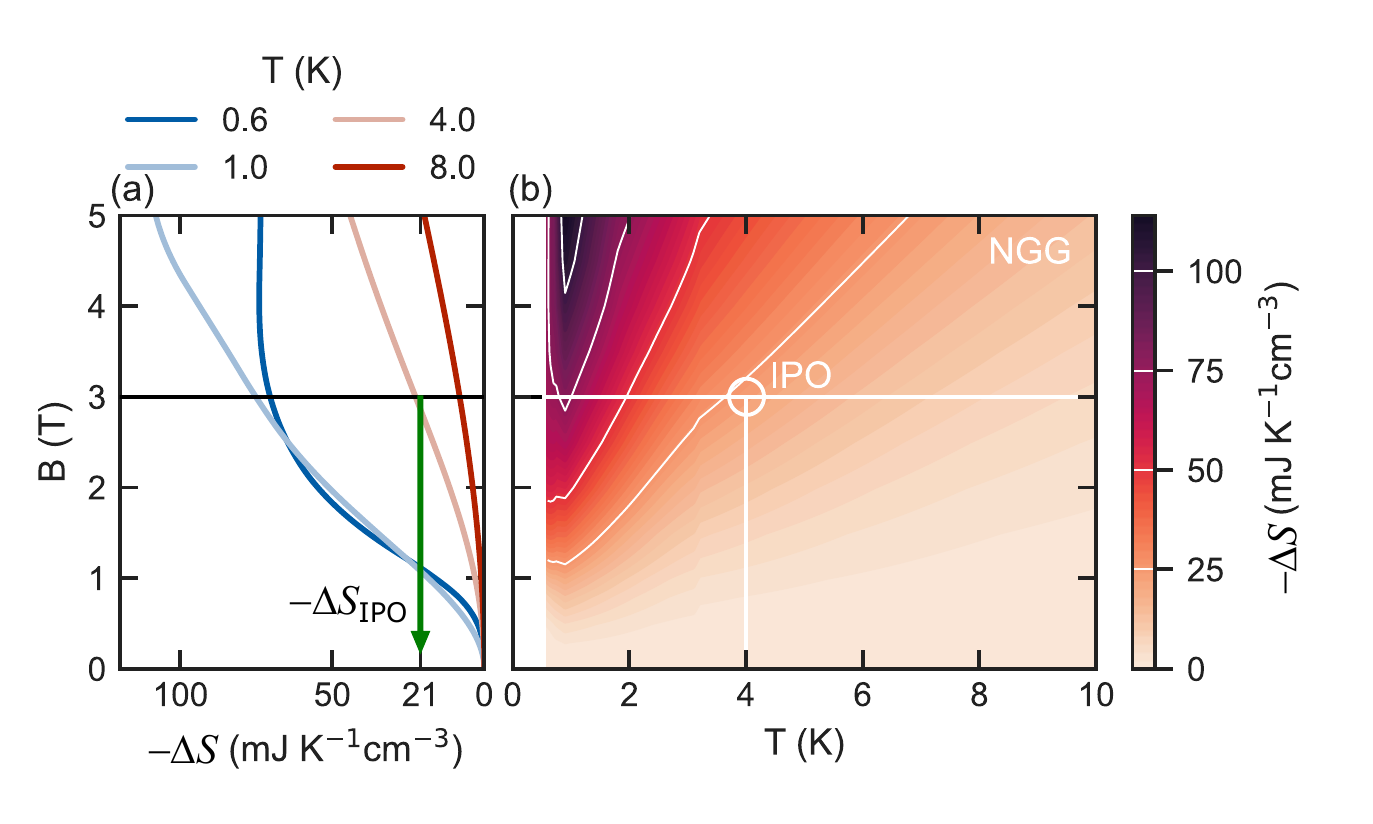}
	\caption{ Figure of merit for the MCE in Nd$_3$Ga$_5$O$_{12}$. A peak in the MCE is reached at 1 K, where it reaches the maximum value of $\Delta S$ = \SI{113}{mJ~K^{-1}cm^{-3}}, as given in Table~\ref{tab:mce_data}. { \bl The IPO is marked in both figures. } }
	\label{fig:ngg_contour}
\end{figure}
\begin{figure}
        \centering
        \includegraphics[width=1.05\columnwidth,clip]{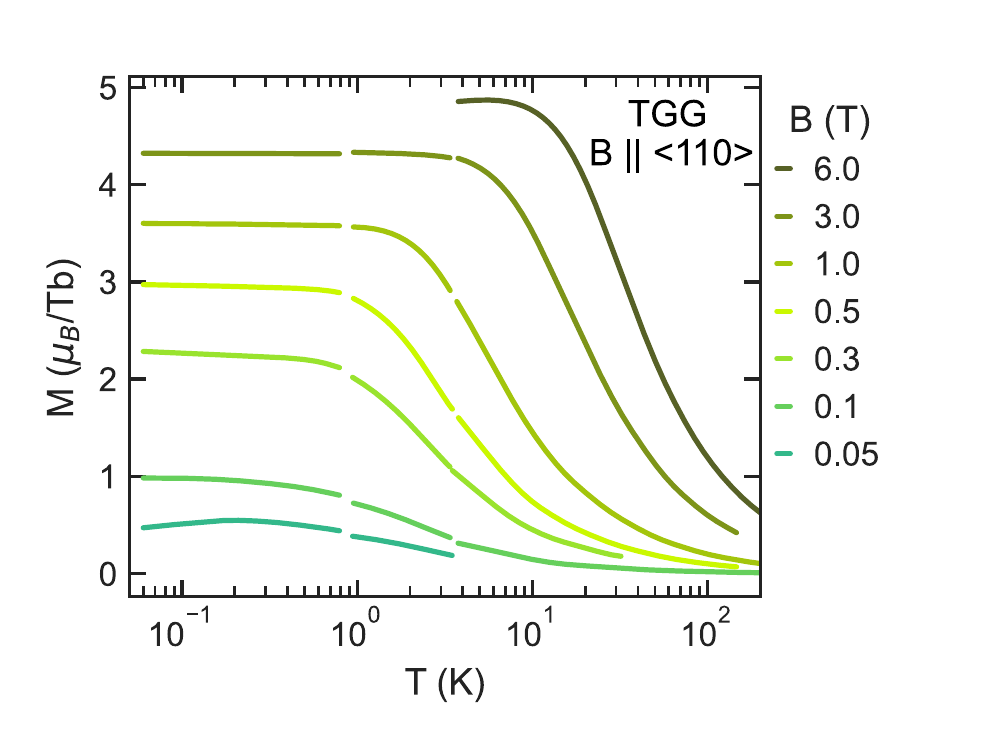}
        \caption{Overview of the temperature scans for TGG obtained in the VSM and VCM. The magnetization per terbium ion is plotted against temperature. Data is recorded from \SI{300}{K} to \SI{60}{mK} under applied magnetic fields between \SI{50}{mT} and \SI{6}{T}. Only data recorded under FH-FC are shown. The magnetic field is applied along the [110] direction.}
        \label{fig:tgg_mag_fullrange}
\end{figure}
    
\begin{figure}
	\centering
	\includegraphics[width=1.05\columnwidth,clip]{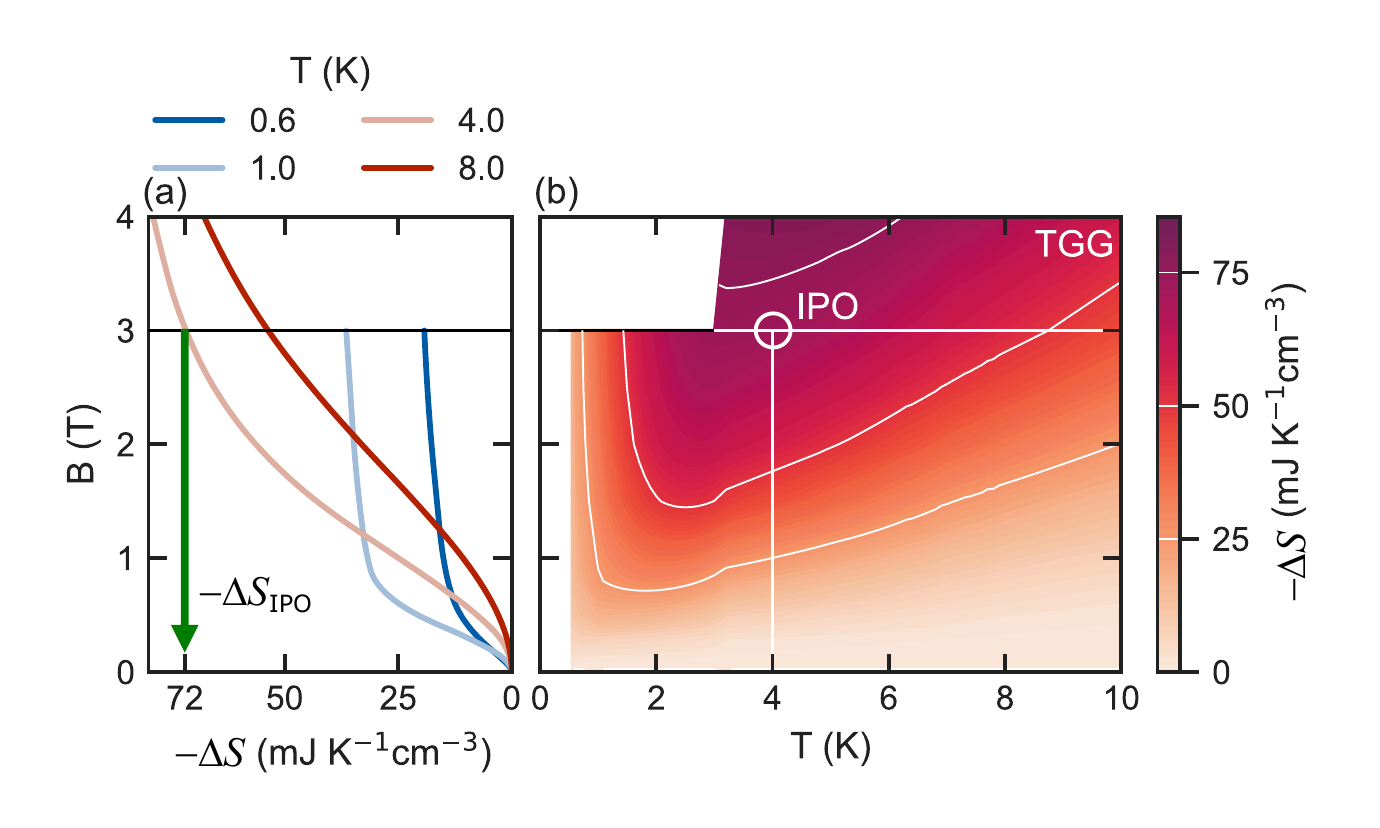}
	\caption{ Figure of merit for the magnetocaloric effect in Tb$_3$Ga$_5$O$_{12}$. A peak in the MCE is reached at approximately \SI{3}{K} and \SI{4}{T}, with an overall maximum value of $\Delta S$~=~\SI{80}{mJ~K^{-1}cm^{-3}}. {\ch Thus,} at the IPO $\Delta S$ nearly reaches its maximum value.  }
	\label{fig:tgg_contour}
\end{figure}

\begin{table*}[t]
        \centering
        \begin{tabular}{cccccccccccc}    
        \hline
        Material, Atom &  g$_{J}$ & $J$ & g$_{J}J$ & $\xi$ & $M_\mathrm{sat}$ & $T_\mathrm{min}$  & $\Delta S_\mathrm{IPO}$ at \SI{4}{K}/\SI{3}{T} & $\mathrm{\Delta S_{max}}$ &  $\delta \mathrm{T_{FWHM}}$ & $S_{RCP,\mathrm{max}}$ \\
 
 &  &  &  \si{\micro_B} &  & \si{\micro_B}/R & \si{mK} & \si{mJ/cm^3 K} & \si{mJ/cm^3 K} & \si{K}  & \si{mJ /cm^3} \\ \hline
        GGG, Gd$^{3+}$  & 2  & 7/2 & 7 & 15.75      & 7.1 {\ch $\pm$ 0.1}             & 800               & 122 {\ch $\pm$ 4}        & 250 {\ch $\pm$ 8} & 9.3 {\ch $\pm$ 0.1} &  2320 {\ch $\pm$ 70} \\
        DGG, Dy$^{3+}$ & 4/3 & 15/2 & 10 & 7.08     & 6.2 {\ch $\pm$ 0.1}           &   $373^{\mathrm{a}}$    &     90 {\ch $\pm$    3}  &  106 {\ch $\pm$ 3} & {\chk 8.9} {\ch $\pm$ 0.2} & {\chk 940} $\pm$ 30  \\
        TGG, Tb$^{3+}$ & 3/2  & 6 & 9 & 10. 5     & $4.75^{\mathrm{b}}$ {\ch $\pm$ 0.05}       & $340^{\mathrm{c}}$    & 72 {\ch $\pm$ 2 }         & 80 {\ch $\pm$ 2}  & 10.1 {\ch $\pm$ 0.1} & 810 {\ch $\pm$ 20}  \\
        NGG, Nd$^{3+}$  & 8/11  & 9/2 & 3.27 & 1.84    & $1.3^{\mathrm{b}}$ {\ch $\pm$ 0.07}          & $640^{\mathrm{c}}$    & 21 {\ch$\pm$ 3 }        & 113 {\ch $\pm$ 4} & 2.8 {\ch $\pm$ 0.05} & 316 {\ch$\pm$ 12}  \\
         \hline 
        \end{tabular}   
        \caption{Summary of the magnetocaloric properties of the four $R_3$Ga$_5$O$_{12}$ compounds observed experimentally . The quantities listed are the saturation magnetization $M_\mathrm{sat}$, the minimal ADR temperature $T_\mathrm{min}$, the MCE ($\Delta S$) at the IPO (\SI{4}{K}, \SI{3}{T}), the maximum MCE in the observed region $\Delta S_\mathrm{max}$, the temperature span $T_\mathrm{FWHM}$ (the temperature interval where the material has its maximum cooling power), {\ch and} the maximum observed value of the $S_{RCP,\mathrm{max}}$. {\ch For} the different rare-earth atoms, the Lande-factor g$_J$, azimuthal quantum number $J$, theoretical maximum saturation magnetization g$_{J}J$, and the \mbox{de Gennes} factor $\xi = (g-1)^2 J(J+1)$ {\ch are given}. ($^{\mathrm{a}}$Antiferromagnetic ordering, information from \cite{filippi1977specific}. $^{\mathrm{b}}$Maximal magnetic moment under the magnetic fields aplied in this study; the literature describes larger values at much higher fields. \cite{wawrzynczak2019magnetic,low2013magnetization,nekvasil1974anisotropy} $^{\mathrm{c}}$Estimated from the upper boundary of the hysteretic temperature range.)}
        \label{tab:mce_data}
\end{table*}

\subsection{Nd$_{3}$Ga$_{5}$O$_{12}$ (NGG)}
Fig.~\ref{fig:ngg_mag_fullrange} shows the magnetization as a function of temperature recorded for NGG across the full temperature range from \SI{300}{K} down to \SI{0.06}{K}. {\ch As before,} these data are used to calculate the magnetocaloric effect. From room temperature down to around \SI{1}{K}, the experimental data exhibit a smooth nonhysteretic increase of the magnetization as expected for a paramagnet. Below \SI{1}{K}, the qualitative behavior is changed. Notably, hysteresis may be observed around \SI{4}{K}, suggestive of a transition to an antiferromagnetic phase.

The results of the evaluation of the MCE for NGG is shown in Fig.~\ref{fig:ngg_contour}. { \bl In the parameter range studied, NGG reaches a maximum magnetocaloric effect $\Delta S_{\mathrm{max}}$ of \SI{113}{mJ~K^{-1}cm^{-3}} at $B$~=~\SI{5}{T} and $T$~=~\SI{1}{K}. At lower $T$, the MCE decreases with decreasing temperature. The MCE values drop off very rapidly as the temperature is increased above \SI{1}{K}.  At the IPO, the change in entropy amounts to \SI{21}{mJ~K^{-1}cm^{-3}}, far below the value of DGG and only a fraction of the maximum value of the MCE. The $RCP$ only attains only a value of \SI{319}{mJ/cm^{3}}}. Large entropy changes in this material can only be achieved at temperatures below \SI{3}{K}. Together with the transition {\ch occuring} at \SI{0.6}{K}, this greatly limits the usable parameter range for NGG.



\subsection{Tb$_3$Ga$_5$O$_{12}$ (TGG)}

The magnetization recorded in TGG is shown in Fig.~\ref{fig:tgg_mag_fullrange}. {\pt All the lines of this plot are obtained by a FH-FC protocol.} The magnetization reaches a maximum of \SI{4.75}{\micro _B/Tb}. Below a temperature of \SI{1}{K}, the magnetization is saturated.  Other measurements showed changes in the magnetization below \SI{0.33}{K}, which may be interpreted in terms of phase transitions. {\pt This coincides with a previously reported transition temperature of approximately \SI{0.25}{K} in the specific heat at zero-field} \cite{hamman:jpa-00207421}. 

In stark contrast to NGG, the MCE in TGG shown in Fig.~\ref{fig:tgg_contour}, is already significantly higher at \SI{10}{K} and an applied field of \SI{4}{T}. The field dependence of the MCE shows that at \SI{4}{K} the entropy does not change significantly in fields $B>\SI{4}{T}$, where $\Delta S=\SI{80}{mJ~K^{-1}cm^{-3}}$. { \bl For starting conditions of \SI{3}{T}/\SI{4}{K} representing the IPO, TGG may provide a change in entropy of \SI{72}{mJ~K^{-1}cm^{-3}}, almost corresponding to its maximum value.} Unfortunately, the MCE starts to decrease already at \SI{3}{K}, saturating (above $B$=\SI{1}{T}) to \SI{30}{mJ~K^{-1}cm^{-3}} at \SI{1}{K}. For higher temperatures, the decline is slow in comparison to NGG,{ \bl and a higher $RCP$  of $S_{RCP}$ = \SI{751}{mJ/cm^{3}} is attained. }
%
%
\\
\\
\section{Conclusions}
{ \bl Rare-earth garnets intrinsically satisfy many of the criteria expected of refrigerants for sub-liquid-helium temperature ADR applications, namely a large entropy density that can be easily reduced through the MCE at magnetic fields less than \SI{3}{T}, {\ch lack of  ordering processes, vanishingly small} magnetic and thermal hysteresis, high chemical stability, non-corrosiveness, high electrical resistivity, and ease of material and device fabrication in a scalable manner. For the four members of the $R_3$Ga$_5$O$_{12}$ family that appear most promising, we reported data needed to critically evaluate their suitability as ADR refrigerants.}

{\bl Amongst the common MCE materials useful within the temperature range 0.3 to \SI{4.2}{K}, GGG has the highest entropy density with $S$=\SI{362}{mJ~K^{-1}cm^{-3}}, {\ch reflecting an excellent MCE cooling capacity}. In addition, it has the combined advantages of being  easy-to-handle and {\ch readily available commercially.} However, GGG requires large fields at \SI{4}{K} ({$>$}~\SI{4}{T}) to access a significant portion of its entropy. It is {\ch also} limited by the fact that its cooling power drops strongly below $T_\mathrm{min}$~=~\SI{0.8}{K}. At the IPO, GGG can only access 50\% of its maximal entropy change (see Table \ref{tab:mce_data}). }


{ \bl DGG has a significantly lower entropy density than GGG, but {\chk a large RCP of $S_{RCP}$ $\sim$ \SI{940}{mJ/cm^{3}}.} At \SI{4}{K} its MCE value is already saturated under an applied field of \SI{4}{T}. At the IPO, the entropy change reaches $\Delta S_\mathrm{IPO}=\SI{90}{mJ~K^{-1}cm^{-3}}$, representing almost its maximum value of $\Delta S_\mathrm{max}=\SI{106}{mJ~K^{-1}cm^{-3}}$}. Moreover, the minimum temperature attainable by demagnetization with DGG is \SI{0.373}{K}, matching the desired range of operation for the applications considered here. 

 { \bl While single crystals of NGG exhibit significant values of the MCE for small fields at temperatures well below 4K, outside of this range its MCE values are comparatively small. NGG reaches only $\Delta S_\mathrm{IPO}=\SI{21}{mJ~K^{-1}cm^{-3}}$,  less than 20\% of its maximum value $\Delta S_\mathrm{max}=\SI{113}{mJ~K^{-1}cm^{-3}}$. Considering also the ordering temperature of about \SI{0.6}{K}, the useful temperature range of NGG is rather limited for the ADR applications under discussion here.}

{ \bl The analysis of our measurements for single-crystal TGG showed a significant entropy change of $\Delta S_\mathrm{IPO}$ = \SI{72}{mJ~K^{-1}cm^{-3}}. However, similar to GGG, it loses a substantial amount of entropy and hence cooling capacity for temperatures below \SI{1}{K}, saturating at a maximum value of $\Delta S$ = \SI{30}{mJ~K^{-1}cm^{-3}} for B$=$ \SI{1}{T}, when higher fields exhibit no substantial effect on the entropy unless the temperature is increased. Thus, neither NGG or TGG are particularly well suited for typical ADR systems \cite{kiutra}, where the first stage of refrigeration is based on a starting temperature of about \SI{4}{K}.}

{\bl In summary, compared to GGG, DGG shows the highest promise for increasing the efficiency of current low temperature ADR systems by virtue of both its minimal accessible temperature of \SI{0.373}{K} and its substantial MCE value which is accessible with lower applied magnetic fields and over a wider temperature range than in the case of GGG. The only remaining hurdle for utilizing DGG in such contexts is finding more efficient ways to produce it in the quantities required as well as sufficient quality. }

\section{Acknowledgments}
{ \bl We thank A. Bauer and S. Mayr for assistance with the experiments and fruitful discussions. This study has been funded by the Deutsche Forschungsgemeinschaft (DFG, German Research Foundation) under TRR80 (From Electronic Correlations to Functionality, Project No.\ 107745057, Projects E1 and F2), SPP2137 (Skyrmionics, Project No.\ 403191981, Grant PF393/19), and the excellence cluster MCQST under Germany's Excellence Strategy EXC-2111 (Project No.\ 390814868). Financial support by the German Bundesministerium für Bildung und Forschung (BMBF, Federal Ministry for Education and Research) through the Project No.\ 13N15021 ('MARQUAND - Magnetische Dauerkühlung für Quantendetektoren'), Project No.\ 05K19W05 ('Resonante Longitudinale MIASANS Spin-Echo Spektroskopie an RESEDA'), and the European Research Council (ERC) through Advanced Grants No.\ 291079 (TOPFIT) and No.\ 788031 (ExQuiSid) is gratefully acknowledged.  }

\bibliography{Garnets_ref}
\end{document}